\documentclass[aps,twocolumn,amsmath,amssymb]{revtex4}

\usepackage[latin1]{inputenc}
\usepackage{bm}
\usepackage{xcolor}
\usepackage{tikz}
\usepackage{graphicx}
\usepackage{subfig}
\usetikzlibrary{trees,arrows,decorations.pathmorphing}

\newcommand{\la}{\langle}
\newcommand{\ra}{\rangle}
\newcommand{\adj}{^{\dagger}}
\newcommand{\mb}{\mathbf}

\newcommand{\mx}{\mathbf{x}}
\newcommand{\mr}{\mathbf{r}}
\newcommand{\beq}{\begin{equation}}
\newcommand{\eeq}{\end{equation}}

\begin{document}

\title{Electronic excitation energies of molecular systems from the Bethe-Salpeter equation: Example of the H$_2$ molecule}
\author{Elisa Rebolini}\email{rebolini@lct.jussieu.fr}
\author{Julien Toulouse}\email{julien.toulouse@upmc.fr}
\author{Andreas Savin}\email{savin@lct.jussieu.fr}
\affiliation{Laboratoire de Chimie Théorique, Université Pierre et Marie Curie and CNRS, 75005 Paris, France}

\date{\today}
\begin{abstract}
We review the Bethe-Salpeter equation (BSE) approach to the calculation of electronic excitation energies of molecular systems. We recall the general Green's function many-theory formalism and give the working equations of the BSE approach within the static GW approximation with and without spin adaptation in an orbital basis. We apply the method to the pedagogical example of the H$_2$ molecule in a minimal basis, testing the effects of the choice of the starting one-particle Green's function. Using the non-interacting one-particle Green's function leads to incorrect energy curves for the first singlet and triplet excited states in the dissociation limit. Starting from the exact one-particle Green's function leads to a qualitatively correct energy curve for the first singlet excited state, but still an incorrect energy curve for the triplet excited state. Using the exact one-particle Green's function in the BSE approach within the static GW approximation also leads to a number of additional excitations, all of them being spurious except for one which can be identified as a double excitation corresponding to the second singlet excited state.
\end{abstract}

\maketitle

\section{Introduction}

Time-dependent density-functional theory (TDDFT)~\cite{RunGro-PRL-84} within the linear response formalism~\cite{GroKoh-PRL-85,Cas-INC-95,PetGosGro-PRL-96} is nowadays the most widely used approach to the calculation of electronic excitation energies of molecules and solids. Applied within the adiabatic approximation and with the usual local or semilocal density functionals, TDDFT gives indeed in many cases excitation energies with reasonable accuracy and low computational cost. However, several serious limitations of these approximations are known, e.g. for molecules: too low charge-transfer excitation energies~\cite{DreWeiHea-JCP-03}, lack of double excitations~\cite{MaiZhaCavBur-JCP-04}, and wrong behavior of the excited-state surface along a bond-breaking coordinate (see, e.g., Ref.~\onlinecite{GriGisGorBae-JCP-00}). Several remedies to these problems are actively being explored, including: long-range corrected TDDFT~\cite{TawTsuYanYanHir-JCP-04,KroSteRefBae-JCTC-12} which improves charge-transfer excitation energies, dressed TDDFT~\cite{MaiZhaCavBur-JCP-04,Cas-JCP-05,HuiCas-ARX-10} which includes double excitations, and time-dependent density-matrix functional theory (TDDMFT)~\cite{PerGriBae-PRA-07,PerGieGriBae-JCP-07,GieBaeGri-PRL-08,GiePerGriBae-JCP-09,Per-JCP-12} which tries to address all these problems.

In the condensed-matter physics community, the Bethe-Salpeter equation (BSE) applied within the GW approximation (see, e.g., Refs.~\onlinecite{Str-RNC-88,RohLou-PRB-00,OniReiRub-RMP-02}) is often considered as the most successful approach to overcome the limitations of TDDFT. Although it has been often used to describe excitons (bound electron-hole pair) in periodic systems, it is also increasingly applied to calculations of excitation energies in finite molecular systems~\cite{Roh-IJQC-00,GroRohMitLouCoh-PRL-01,TiaChe-SSC-05,HahSchBec-PRB-05,TiaChe-PRB-06,TiaKenHooReb-JCP-08,GruMarGon-NL-09,MaRohMol-PRB-09,MaRohMol-JCTC-10,RocLuGal-JCP-10,GruMarGon-CMS-11,BlaAtt-APL-11}. In particular, the BSE approach is believed to give accurate charge-transfer excitation energies in molecules~\cite{RocLuGal-JCP-10,BlaAtt-APL-11}, and when used with a frequency-dependent kernel it is in principle capable of describing double excitations~\cite{RomSanBerSotMolReiOni-JCP-09,SanRomOniMar-JCP-11}. 

In this work, we examine the merits of the BSE approach for calculating excitation energies of the prototype system of quantum chemistry, the H$_2$ molecule. The paper is organized as follows. In Sec.~\ref{sec:review}, we give a review of Green's function many-body theory. In Sec.~\ref{sec:finitebasis}, we give the general working equations for a BSE calculation within the static GW approximation in a finite spin-orbital basis, and the corresponding spin-adapted expressions for closed-shell systems. In Sec.~\ref{sec:h2}, we apply the equations to the H$_2$ molecule in a minimal basis and discuss the possibility of obtaining correct spin-singlet and spin-triplet excited-state energy curves as a function of the internuclear distance. Sec.~\ref{sec:conclusion} contains our conclusions.

\section{Review of Green's function many-body theory}
\label{sec:review}

We start by giving a brief review of Green's function many-body theory for calculating excitation energies. For more details, see e.g. Refs.~\onlinecite{Str-RNC-88,OniReiRub-RMP-02,Bru-THESIS-05}.

\subsection{One-particle Green's function}

Let $|N\ra$ be the normalized ground-state wave function for a system of $N$ electrons described by the Hamiltonian $\hat{H}$. 
The time-ordered one-particle equilibrium Green's function is defined as
\begin{equation}
\begin{split}
iG(1,2) = & \la N | \hat{T} [\hat{\Psi}(1) \hat{\Psi}\adj(2) ]| N \ra \\
		= & \theta(t_1-t_2)\la N | \hat{\Psi}(1) \hat{\Psi}\adj(2) | N \ra \\
		&	-\theta(t_2-t_1)\la N | \hat{\Psi}\adj(2) \hat{\Psi}(1)  | N \ra.
\end{split}
\end{equation}
Index 1 stands for space, spin and time coordinates $(\mr_1,\sigma_1,t_1)=(\mx_1,t_1)$.
$\hat{T}$ is the Wick time-ordering operator which orders the operators with  larger times on the left and $\theta$ is the Heaviside step function.
The whole time-dependence is contained in 
$\hat{\Psi}(1) = e^{i\hat{H}t_1}\hat{\Psi}(\mx_1)e^{-i\hat{H}t_1}$ and $\hat{\Psi}\adj(2)= e^{i\hat{H}t_2}\hat{\Psi}\adj(\mx_2)e^{-i\hat{H}t_2}$, the annihilation and creation field operators in the Heisenberg representation.

In the absence of external potential, the system is invariant under time translation, therefore the Green's function depends only on $\tau = t_1-t_2$. By introducing the closure relation for excited states with $N-1$ or $N+1$ particles, one can get
\begin{equation}
	\begin{split}
	iG (&\mx_1,\mx_2;\tau) = \\
		& \theta(\tau) \sum_A  \la N | \hat{\psi}(\mx_1) | N+1,A \ra \la N +1,A|\hat{\psi}\adj(\mx_2) | N \ra \\
		& \hspace*{1cm} \times e^{-i(E_{N+1,A}-E_N)\tau}  \\
		& - \theta(-\tau) \sum_I \la N | \hat{\psi}\adj(\mx_2)  | N-1,I\ra  \la N-1,I|\hat{\psi}(\mx_1)| N \ra \\
		& \hspace*{1cm} \times e^{-i(E_{N}-E_{N-1,I})\tau},
	\end{split}	
\label{Gt}
\end{equation}
where $E_N$, $E_{N+1,A}$ and $E_{N-1,I}$ are the energies of the ground state $|N\ra$, of the $A$-th excited state with $N+1$ particles $|N+1,A \ra$ and of the $I$-th excited state with $N-1$ particles $|N-1,I\ra$, respectively.
The Lehmann representation of the one-particle Green's function is obtained by Fourier transform
\begin{equation}
G(\mx_1,\mx_2;\omega) = \sum_{A} \dfrac{f_A(\mx_1)f_A^*(\mx_2)}{\omega - {\cal E}_A + i0^+}
+\sum_I \dfrac{f_I(\mx_1) f_I^*(\mx_2)}{\omega - {\cal E}_I - i0^+},
\label{GLehmann}
\end{equation}
where $f_A(\mx) = \la N|\hat{\psi}(\mx)|N+1,A\ra$ and $f_I(\mx) = \la N-1,I | \hat{\psi}(\mx)|N\ra$ are the Dyson orbitals, and ${\cal E}_A = E_{N+1,A}-E_N$ and ${\cal E}_I = E_N-E_{N-1,I}$ are minus the electron affinities and ionization energies, respectively.

\subsection{Two-particle Green's function}

The time-ordered two-particle Green's function is defined as
\begin{equation}
i^2 G_2(1,2;1',2') = \la N | \hat{T} [\hat{\Psi}(1)\hat{\Psi}(2) \hat{\Psi}\adj(2') \hat{\Psi}\adj(1') ]| N \ra .
\end{equation}
Depending on the time ordering, it describes the propagation of a pair of holes, of electrons or of a hole and an electron.
In the case of optical absorption, one is only interested in the propagation of a hole-electron  pair.

Let $\chi$ be the 4-point polarizability, 
\begin{equation}
\chi(1,2;1',2') = iG_2(1,2;1',2') -i G(1,1')G(2,2').
\end{equation}
It describes the coupled motion of two particles minus the motion of the independent ones.
When the times are appropriately ordered, the 4-point polarizability reduces to the linear response function
\begin{equation}
\chi(\mx_1,\mx_2;\mx_1',\mx_2';\tau) = \chi(\mx_1,t_1,\mx_2,t_2;\mx_1',t_1^+,\mx_2',t_2^+)
\end{equation}
where $t_1^+ = t_1 + 0^+$.
The Lehmann representation of the response function explicitly gives the excitation energies as poles in $\omega$,
\begin{equation}
\begin{split}
\chi&(\mx_1,\mx_2;\mx_1',\mx_2';\omega) = \\
& \sum_{K\neq 0} \dfrac{\la N | \hat{\Psi}\adj(\mx_1') \hat{\Psi}(\mx_1) | N,K \ra \la N ,K| \hat{\Psi}\adj(\mx_2') \hat{\Psi}(\mx_2) | N \ra}{\omega - (E_{N,K} - E_N) + i0^+} \\
	& - \sum_{K \neq 0} \dfrac{\la N | \hat{\Psi}\adj(\mx_2') \hat{\Psi}(\mx_2) | N,K \ra \la N ,K| \hat{\Psi}\adj(\mx_1') \hat{\Psi}(\mx_1) | N \ra}{\omega + (E_{N,K} - E_N) - i0^+} ,
\end{split}
\end{equation}
where $|N,K\ra$ is the $K$-th excited state with $N$ particles of energy $E_{N,K}$. The ground state $|N,0\ra = |N \ra $ is excluded from the sum. It is also useful to define the independent-particle (IP) polarizability $\chi_{IP}(1,2;1',2') = -i G(1,2') G(2,1')$. Its Lehmann representation is easily obtained by calculating $\chi_{IP}(\mx_1,\mx_2;\mx_1',\mx_2';\tau) = -i G(\mx_1,\mx_2';\tau) G(\mx_2,\mx_1';-\tau)$ with equation~(\ref{Gt}) and taking the Fourier transform
\begin{eqnarray}
\chi_{IP}(\mx_1,\mx_2;\mx_1',\mx_2';\omega) = \sum_{IA} 
\dfrac{f_I^*(\mx_1')f_A(\mx_1)f_A^*(\mx_2')f_I(\mx_2) }
{\omega -({\cal E}_A -{\cal E}_I) +i0^+}
\nonumber\\
-
\sum_{IA} \dfrac{f_I^*(\mx_2')f_A(\mx_2)f_A^*(\mx_1')f_I(\mx_1) }
{\omega +({\cal E}_A -{\cal E}_I) -i0^+}.
\label{ChiIPLehmann}
\end{eqnarray}

In practice, the one-particle and two-particle Green's function can be calculated with equations of motion.

\subsection{Dyson equation}
To make easier the connection with expressions in a finite spin-orbital basis, we systematically use 4-point indexes for all the two-electron quantities.
The starting point is therefore a fully non-local time-dependent Hamiltonian,
\begin{equation}
\begin{split}
\hat{H}&(t_1) = \int d \mx_1 d1'  \hat{\Psi}\adj(1)h(1,1') \hat{\Psi}(1') \\
 	& + \dfrac{1}{2}\int d \mx_1 d2 d1' d2' \hat{\Psi}\adj(1) \hat{\Psi}\adj(2)v(1,2;1',2') \hat{\Psi}(1') \hat{\Psi}(2'),
\end{split}
\end{equation}
where $v(1,2;1',2')= v_{ee}(|\mr_1-\mr_2|) \delta(t_1,t_2)\delta(1,1') \delta(2,2')$ is the spin-independent instantaneous Coulomb electron-electron interaction and $ h(1,1')$ is the one-electron Hamiltonian which contains the electron kinetic operator and the nuclei-electron interaction $V_{ne}$,
\begin{equation}
h(1,1') = -\delta(1,1') \dfrac{\nabla_1^2}{2} + \delta(1,1') V_{ne}(\mr_1).
\end{equation}
Using the equations of motion for the Heisenberg creation and annihilation operators in the expression of the derivative of $G$ with respect to time~\cite{Str-RNC-88}, one can obtain the following equation, 
\begin{equation}\label{eq:motion}
\begin{split}
i&\int d3 \delta(1,3)\dfrac{\partial}{\partial t_1}G(3,2) - \int d3 h(1,3) G(3,2) \\
&+ i \int d3d1'd3' v(1,3;1',3') G_2(1',3'^+;2,3^{++}) = \delta(1,2),
\end{split}
\end{equation}
where $^{++}$ stands for $t_3^+ +0^+$.
A whole series of equations can be derived for the Green's functions, relating
the one-particle Green's function to the two-particle Green's function, the two-particle one to the three-particle one, etc. But solving this set of equations is not wanted.

To avoid this, one can use the Schwinger derivative technique.
Introducing an external time-dependent potential $U(1,1') = U(\mx_1,\mx_1',t_1)\delta(t_1,t_1')$, one can express the two-particle Green's function in terms of the one-particle one and of its derivative with respect to $U$, evaluated at $U=0$,
\begin{equation}
\dfrac{\delta G(1,2)}{\delta U(3,4)} = - G_2(1,4;2,3) + G(1,2)G(4,3).
\end{equation}
Using this relation in equation (\ref{eq:motion}), one can get 
\begin{equation}\label{eq:gen_dyson}
\begin{split}
\int  d3 & \left[ i\delta(1,3)\dfrac{\partial}{\partial t_1} 
 -  h(1,3) \right] G(3,2) \\
 & - \int d3 \, \Sigma_{Hxc}(1,3)G(3,2) = \delta(1,2),
\end{split}
\end{equation}
where $\Sigma_{Hxc}(1,2) $ is the Hartree-exchange-correlation self-energy which takes into account all the two-particle effects.
It can be decomposed into a Hartree contribution
\begin{equation}
\Sigma_H(1,2) = -i \int d3d3' v(1,3;2,3') G(3'^+,3^{++}),
\end{equation} 
and an exchange-correlation one
\begin{equation}
\begin{split}
\Sigma&_{xc}(1,2) = \\
& i\int d3 d1' d3' d4 v(1,3;1',3') \dfrac{\delta G(1',4)}{\delta U(3^{++},3'^+)}G^{-1}(4,2).
\end{split}
\end{equation} 
One can define a Green's function $G_h$ which shows no two-particle effects and therefore follows the equation of motion 
\begin{equation}
\int d3 \left[ i\delta(1,3)\dfrac{\partial }{\partial t_1} -h(1,3) \right] G_h(3,2)  = \delta(1,2).
\end{equation}
Using this relation in equation (\ref{eq:gen_dyson}), one finally gets the Dyson equation for the one-particle Green's function,
\begin{equation}
\int d3\left[ G_h^{-1}(1,3) - \Sigma_{Hxc}(1,3) \right]G(3,2) = \delta(1,2).
\end{equation}
This equation is also often used under the forms
\begin{equation}\label{eq:dyson_SCF}
G(1,2) = G_h(1,2) + \int d3d4 G_h(1,3) \Sigma_{Hxc}(3,4) G(4,2),
\end{equation}
or
\begin{equation}\label{eq:dyson_-1}
G^{-1}(1,2) = G_h^{-1}(1,2) - \Sigma_{Hxc}(1,2).
\end{equation}

\subsection{Bethe-Salpeter Equation}
Starting from the Dyson equation (\ref{eq:dyson_-1}), and taking the derivative with respect to $G$, one can get the so-called Bethe-Salpeter equation (see, e.g., Ref.~\onlinecite{TouZhuAngSav-PRA-10})
\begin{equation}
\chi^{-1}(1,2;1',2') = \chi_{IP}^{-1}(1,2;1',2') -\Xi_{Hxc}(1,2;1',2'),
\end{equation}
or
\begin{equation}
\begin{split}
\chi&(1,2;1',2') = \chi_{IP}(1,2;1',2') \\
&+ \int d3456 \chi_{IP}(1,4;1',3) \Xi_{Hxc}(3,6;4,5) \chi(5,2;6,2'),
\end{split}
\end{equation}
where $\Xi_{Hxc}$ is the Hartree-exchange-correlation Bethe-Salpeter kernel, defined as
\begin{equation}
\Xi_{Hxc}(3,6;4,5) = i \dfrac{\delta \Sigma_{Hxc}(3,4)}{\delta G(5,6)}.
\end{equation}

\subsection{Hedin's equations}
We now have equations of motion for the one- and two-particle Green's functions.
They depend on the Hartree-exchange-correlation self-energy.
Only its Hartree part is known exactly.
A practical way of calculating its exchange-correlation part is needed. 
Hedin proposed a scheme which yields to a set of coupled equations and allows in principle for the calculation of the exact self-energy~\cite{hedin_new_1965}.
This scheme can be seen as a perturbation theory in terms of the screened interaction $W$ instead of the bare Coulomb interaction $v$.
We show a generalization of this derivation for the case of a non-local potential.

Let $V(5,6) = U(5,6) - i\int d3d3'v(5,3;6,3')G(3',3^+)$ be the non-local classical potential.
Using the chain rule in the exchange-correlation self-energy, we get: 
\begin{equation}\label{eq:hedin}
\begin{split}
\Sigma_{xc}(1,2)	
& =- i\int d3 d1' d3' d4 d5 d6 v(1,3;1',3') G(1',4) \\
& \hspace*{1cm}\times \dfrac{\delta G^{-1}(4,2)}{\delta V(5,6)}\dfrac{\delta V(5,6)}{\delta U(3^{++},3'^+)} \\
& =  i\int d3 d1' d3' d4 d5 d6 v(1,3;1',3') G(1',4) \\
& \hspace*{1cm}\times  \tilde{\Gamma}(4,6;2,5)\epsilon^{-1}(5,3';6,3^+).
\end{split}
\end{equation}
where  the inverse dielectric function $\epsilon^{-1}$ which screens the bare Coulomb interaction $v$ and the irreducible vertex function $\tilde{\Gamma}$ are defined by 
\begin{equation}
\begin{split}
\epsilon^{-1}&(1,2;3,4) = \dfrac{\delta V(1,3)}{\delta U (4,2)} \\ 
& \text{and \hspace*{0.1cm}}
\tilde{\Gamma}(1,2;3,4) = - \dfrac{\delta G^{-1}(1,3)}{\delta V(4,2)}.
\end{split}
\end{equation}
 We can therefore define a dynamically screened potential 
\begin{equation}\label{eq:hedin_W}
\begin{split}
W(1,2;1',2') &= \int d3d3' \epsilon^{-1}(1,3;1',3'^+) v(2,3';2',3) \\
&= \int d3d3' \epsilon^{-1}(1,3;1',3'^+) v(3',2;3,2'),
\end{split}
\end{equation}
where the symmetry of the Coulomb interaction $v$ has been used, and we get the expression of the exchange-correlation self-energy, 
\begin{equation}
\begin{split}
\Sigma_{xc}&(1,2)=\\
& i \int d1'd3d3'd4 G(1',4)\tilde{\Gamma}(4,3',2,3) W(3,1;3',1').
\end{split}
\end{equation}
We still need to express the dielectric function and the irreducible vertex function without the use of $V$ and $U$.
To achieve this, we define the irreducible polarizability 
$\tilde{\chi}(1,2;3,4) = -i \delta G(1,3) / \delta V(4,2)$,
 which, with the properties of the inverse and the definition of the vertex correction, can be rewritten as
\begin{equation}
\tilde{\chi}(1,2;3,4) = -i \int d5d5' G(1,5)G(5',3) \tilde{\Gamma}(5,2;5',4).
\end{equation}
Using this relation, one can rewrite the dielectric function as 
\begin{equation}\label{eq:hedin_epsilon}
\begin{split}
\epsilon(1,&2;3,4)= \\
&\delta(1,4)\delta(2,3) -  \int d5d5'v(1,5;3,5') \tilde{\chi}(5',2;5^+,4),
\end{split}
\end{equation}
and the irreducible vertex correction as 
\begin{equation}\label{eq:hedin_Gamma}
\begin{split}
\tilde{\Gamma}(1,&2;3,4) = \\
&\delta(1,4)\delta(2,3) -i\int d5d6 \dfrac{\delta \Sigma_{xc}(1,3)}{\delta G(5,6)}\tilde{\chi}(5,2;6,4).
\end{split}
\end{equation}
We now have a set of five coupled equations (\ref{eq:hedin_W}) to (\ref{eq:hedin_Gamma}) to calculate the self-energy. 
In practice, this set of equations is never solved exactly, approximations are made.

\subsection{Static GW approximation}

We discuss now the static GW approximation which is the most often used approximation in practice in the BSE approach.

In the GW approximation, one takes
$\tilde{\Gamma}(1,2;3,4) = \delta(1,4) \delta(2,3)$.
This simplifies greatly Hedin's equations.
The irreducible polarizability becomes $\tilde{\chi}(1,2;3,4) = -i G(1,4)G(2,3) = \chi_{IP}(1,2;3,4)$ and
the exchange-correlation self-energy becomes
\begin{equation}
\Sigma_{xc}(1,2)  = i \int d1' d3  G(1',3) W(3,1;2,1').
\end{equation}
If the derivative of $W$ with respect to $G$ is further neglected, as usually done, the corresponding Bethe-Salpeter kernel is then
\begin{equation}
\Xi_{Hxc}(1,2;1',2') = v(1,2;1',2')- W(2,1;1',2'),
\end{equation}
where $W$ is obtained from equation (\ref{eq:hedin_W}) and $\epsilon^{-1}$ with equation (\ref{eq:hedin_epsilon}) in which $\tilde{\chi}$ is replaced by $\chi_{IP}$. The Coulomb interaction is instantaneous and the one-particle Green's functions depends only of the time difference, therefore the time dependence of the screened interaction is
\begin{equation}
W(1,2;1',2') = W(\mx_1,\mx_2;\mx_1',\mx_2';\tau) \delta(t_1,t_1')\delta(t_2,t_2'),
\end{equation}
where $\tau = t_1-t_2$.
If one considers the time dependence in $W$, the Fourier transform of the Bethe-Salpeter equation is not straightforward~\cite{RomSanBerSotMolReiOni-JCP-09}.
We will only consider the usual approximation where the screened interaction is static, i.e.,
\begin{equation}
W(1,2;1',2') = W(\mx_1,\mx_2;\mx_1',\mx_2')\delta(t_1,t_1')\delta(t_2,t_2')\delta(t_1,t_2).
\end{equation}
To summarize, the Fourier-space Bethe-Salpeter equation in the static GW approximation writes
\begin{equation}
\begin{split}
\chi^{-1}(\mx_1,&\mx_2;\mx_3,\mx_4;\omega) =\\
& \chi^{-1}_{IP}(\mx_1,\mx_2;\mx_3,\mx_4;\omega) - \Xi_{Hxc}(\mx_1,\mx_2;\mx_3,\mx_4),
\end{split}
\label{BSEGW}
\end{equation}
where the kernel $\Xi_{Hxc}(\mx_1,\mx_2;\mx_3,\mx_4) = v(\mx_1,\mx_2;\mx_3,\mx_4) -W(\mx_2,\mx_1;\mx_3,\mx_4)$ contains the static screened interaction $W$ calculated from
\begin{equation}
\begin{split}
W(\mx_1,&\mx_2;\mx_1',\mx_2') = \\
&\int d\mx_3d\mx_3'  \epsilon^{-1}(\mx_1,\mx_3;\mx_1',\mx_3') v(\mx_3',\mx_2;\mx_3,\mx_2'),
\end{split}
\label{Wstatic}
\end{equation}
and
\begin{equation}
\begin{split}
\epsilon&(\mx_1,\mx_2;\mx_3,\mx_4)=\delta(\mx_1,\mx_4)\delta(\mx_2,\mx_3) \\
&-  \int d\mx_5d\mx_5' v(\mx_1,\mx_5;\mx_3,\mx_5') \chi_{IP}(\mx_5',\mx_2;\mx_5,\mx_4;\omega=0).
\end{split}
\label{epsilonstatic}
\end{equation}
We will refer to the approach of equations~(\ref{BSEGW})-(\ref{epsilonstatic}) as the BSE-GW method. The one-particle Green's function $G$ in $\chi_{IP} = -i GG$ is not yet specified. Different choices can be made. The simplest option is to use a non-interacting Green's function $G_0$ from a Hartree-Fock (HF) or Kohn-Sham (KS) calculation. In this case, $\chi_{IP}=-iG_0 G_0 = \chi_0$ is just the non-interacting HF or KS response function. In the condensed-matter physics literature, the usual recipe is to use $\chi_0$ in equation~(\ref{epsilonstatic}) but an improved $\chi_{IP}$ in equation~(\ref{BSEGW}) from a GW calculation. In the case of H$_2$ in a minimal basis, it is simple enough to use $\chi_{IP}$ constructed with the exact one-particle Green's function $G$. Finally, we note that the dielectric function of equation~(\ref{epsilonstatic}) could be alternatively defined as including the HF exchange in addition to the Coulomb interaction, i.e. $v(\mx_1,\mx_5;\mx_3,\mx_5')\to v(\mx_1,\mx_5;\mx_3,\mx_5') - v(\mx_5,\mx_1;\mx_3,\mx_5')$ (see, e.g., Ref.~\onlinecite{ShiMar-PRB-93}), which removes the ``self-screening error'' for one-electron systems~\cite{RomGuyRei-JCP-09}, but we do not explore this possibility here.

\section{Expressions in a finite orbital basis}
\label{sec:finitebasis}

\subsection{Spin-orbital basis}
In order to solve the Bethe-Salpeter equation for finite systems, all the equations are projected onto an orthonormal spin-orbital basis $\{ \phi_p \}$.
As the equations are 4-point equations relating two-particle quantities, they are in fact projected onto the basis of products of two spin orbitals.
Each matrix element is thus indexed by two double indices. 

We consider the simplest case for which $\chi_{IP} = \chi_0$. The Lehmann representation of $\chi_0$ is 
\begin{equation}
\begin{split}
\chi_0(\mx_1,\mx_2;&\mx_1',\mx_2';\omega) = \\
&\sum_{ia} 
\dfrac{\phi_i^*(\mx_1')\phi_a(\mx_1)\phi_a^*(\mx_2')\phi_i(\mx_2) }
{\omega -(\varepsilon_a - \varepsilon_i) +i0^+}\\
&-\dfrac{\phi_i^*(\mx_2')\phi_a(\mx_2)\phi_a^*(\mx_1')\phi_i(\mx_1) }
{\omega +(\varepsilon_a - \varepsilon_i) -i0^+},
\end{split}
\end{equation}
where $\phi_i$ is the $i$-th occupied spin-orbital of energy $\varepsilon_i$ and $\phi_a$ is the $a$-th virtual spin-orbital of energy $\varepsilon_a$.
One can notice that $\chi_0$ is expanded only on occupied-virtual (ov) and virtual-occupied (vo) products of spin-orbitals. 
The matrix elements of $\chi_0$ are given by
\begin{equation}
\begin{split}
\left[\chi_{0}(\omega)\right]_{pq,rs} =&\int d\mx_1 d\mx_1' d\mx_2 d\mx_2' \phi_p(\mx_1') \phi^*_q(\mx_1)\\
& \times \chi_0(\mx_1,\mx_2;\mx_1',\mx_2';\omega) \phi_r^*(\mx_2) \phi_s(\mx_2').
\end{split}
\end{equation}
The matrix representation of its inverse, in the (ov,vo) subspace, is
\begin{equation}
\bm{\chi}_0^{-1}(\omega)=-\left[
\left(
\begin{array}{cc}
\bm{\Delta \varepsilon} & \bm{0} \\
\bm{0} & \bm{\Delta \varepsilon}
\end{array}
\right)
- \omega
\left(
\begin{array}{cc}
\bm{1} & \bm{0} \\
\bm{0} & \bm{-1}
\end{array}
\right) \right],
\end{equation}
where $\Delta \varepsilon_{ia,jb} = \Delta \varepsilon_{ai,bj} = (\varepsilon_a - \varepsilon_i) \delta_{ij} \delta_{ab}$, where $i,j$ refer to occupied spin-orbitals and $a,b$ to virtual orbitals. The dimension of the matrix is thus $2M_oM_v\times2M_oM_v$ where $M_o$ and $M_v$ are the numbers of occupied and virtual spin orbitals, respectively.
To build the matrix $\bm{\chi}^{-1}$, one then needs to construct the matrix elements of the Bethe-Salpeter kernel $\Xi_{Hxc}$ which are given by
\begin{equation}\label{eq:xi_BSE}
(\Xi_{Hxc})_{pq,rs} = v_{pq,rs} - W_{pr,qs},
\end{equation}
where $v_{pq,rs}=\la qr |ps \ra $ are the usual two-electron integrals, and the matrix elements of $W$ can be obtained from equation~(\ref{Wstatic})
\begin{equation}
\begin{split}
W_{pq,rs}
& = \int d\mx_1 d\mx_1' d\mx_2 d\mx_2' 
\phi_p(\mx_1') \phi^*_q(\mx_1) \\
& \times
W(\mx_1, \mx_2;\mx'_1,\mx'_2)
\phi^*_r(\mx_2) \phi_s(\mx_2')
 \\
& = \int d\mx_1 d\mx_1' d\mx_2 d\mx_2' d\mx_3 d\mx_3'
\phi_p(\mx_1')
\phi^*_q(\mx_1) \\
& \times \epsilon^{-1}(\mx_1,\mx_3;\mx_1',\mx_3')v(\mx_3',\mx_2;\mx_3,\mx_2')	 
\phi^*_r(\mx_2)
\phi_s(\mx_2').
\end{split}
\end{equation}
To decouple the common coordinates in $\epsilon^{-1}$ and $v$, one can introduce two delta functions $\delta(\mx_3,\mx_4)$ and $\delta(\mx_3',\mx_4')$ and use the closure relations $\delta(\mx_3,\mx_4)=\sum_t \phi_t^*(\mx_3)\phi_t(\mx_4)$ and $\delta(\mx_3',\mx_4')=\sum_u \phi_u(\mx_3')\phi_u^*(\mx_4')$.
By doing so, the matrix elements of $v$ and $\epsilon^{-1}$ appear explicitly and we get
\begin{equation}
W_{pq,rs} = \sum_{tu} \epsilon^{-1}_{pq,tu} v_{tu,rs}.
\end{equation}
Similarly, for the dielectric function, we have 
\begin{equation}
\begin{split}
\epsilon_{pq,rs} &= \delta_{pr}\delta_{qs} - \sum_{tu} v_{pq,tu} \left[\chi_{0}(\omega=0)\right]_{tu,rs} \\
&= \delta_{pr}\delta_{qs} -  v_{pq,rs} \left[\chi_{0}(\omega=0)\right]_{rs,rs},
\end{split}
\end{equation}
where the last equality comes from the fact that $\chi_{0}$ has only diagonal elements. It can be seen that the static screened interaction consists of an infinite-order perturbation expansion in the Coulomb interaction, namely using matrix notations,
\begin{equation}
\begin{split}
\bm{W} &= \bm{\epsilon}^{-1} \cdot \bm{v} \\
&= \bm{v} + \bm{v} \cdot \bm{\chi}_{0}(\omega=0) \cdot \bm{v} \\
& \hspace*{0.4cm}+ \bm{v} \cdot \bm{\chi}_{0}(\omega=0) \cdot \bm{v} \cdot \bm{\chi}_{0}(\omega=0) \cdot \bm{v} +...,
\end{split}
\end{equation}
the first term in this expansion corresponding to time-dependent Hartree-Fock (TDHF). The matrix representation of the inverse of the interacting response function, in the (ov,vo) subspace, is then
\begin{equation}
\bm{\chi}^{-1}(\omega)=-\left[
\left(
\begin{array}{cc}
\bm{A} & \bm{B} \\
\bm{B}^* & \bm{A}^*
\end{array}
\right)
- \omega
\left(
\begin{array}{cc}
\bm{1} & \bm{0} \\
\bm{0} & \bm{-1}
\end{array}
\right) \right],
\label{chim1}
\end{equation}
with the matrices
\begin{subequations}
\begin{eqnarray}
A_{ia,jb} = \Delta \varepsilon_{ia,jb} +  v_{ia,jb} - W_{ij,ab},
\end{eqnarray}
\begin{eqnarray}
B_{ia,jb} = v_{ia,bj} - W_{ib,aj}.
\end{eqnarray}
\label{}
\end{subequations}
The block structure of equation~(\ref{chim1}) is a consequence of the symmetry of the Coulomb interaction, $v_{qp,sr}= v_{pq,rs}^*$, and of the static screened interaction, $W_{qs,pr}= W_{pr,qs}^*$. Moreover, the matrix $\bm{A}$ is Hermitian (because $v_{ia,jb} = v_{jb,ia}^*$ and $W_{ij,ab}=W_{ji,ba}^*$) and the matrix $\bm{B}$ is symmetric (because $v_{ia,bj}=v_{jb,ai}$ and $W_{ib,aj}=W_{ja,bi}$). The excitation energies $\omega_n$ are thus found by solving the usual linear-response pseudo-Hermitian eigenvalue equation, just as in TDDFT,
\begin{equation}
\left(
\begin{array}{cc}
\bm{A} & \bm{B} \\
\bm{B}^* & \bm{A}^*
\end{array}
\right) 
\left(
\begin{array}{c}
\bm{X}_n \\
\bm{Y}_n\\
\end{array}
\right) 
= 
\omega_n
\left(
\begin{array}{cc}
\bm{1} & \bm{0} \\
\bm{0} & \bm{-1}
\end{array}
\right) 
\left(
\begin{array}{c}
\bm{X}_n \\
\bm{Y}_n\\
\end{array}
\right),
\label{nonHermeq}
\end{equation}
whose solutions come in pairs: excitation energies $\omega_{n}$ with eigenvectors $\left( \bm{X}_{n}, \bm{Y}_{n} \right)$, and de-excitation energies $-\omega_{n}$ with eigenvectors $\left( \bm{Y}_{n}^*, \bm{X}_{n}^* \right)$. For real spin-orbitals and if $\bm{A}+\bm{B}$ and $\bm{A}-\bm{B}$ are positive definite, the eigenvalues are guaranteed to be real numbers and the pseudo-Hermitian eigenvalue equation~(\ref{nonHermeq}) can be transformed into a half-size symmetric eigenvalue equation~\cite{Cas-INC-95}.

If instead of starting from $\chi_0$, one starts from $\chi_{IP}=-i G G$ with the exact one-particle Green's function $G$, the equations get more complicated since the matrix representation of $\chi_{IP}$ is generally not diagonal and not only has contributions in the (ov,vo) subspace of spin-orbital products but also in the occupied-occupied (oo) and virtual-virtual (vv) subspace of spin-orbital products. The dimension of the matrices thus becomes $M^2\times M^2$ where $M$ is the total number of (occupied and virtual) spin orbitals. In this case, the number of solutions of the response equations is generally higher than the number of single excitations, and in particular double excitations might be obtained even without a frequency-dependent kernel. Spurious excitations can also be found. This is similar to what happens in linear-response TDDMFT~\cite{PerGriBae-PRA-07,PerGieGriBae-JCP-07,GieBaeGri-PRL-08,GiePerGriBae-JCP-09}. We will show this later in the case of H$_2$ in a minimal basis.

\subsection{Spin adaptation}

We give now the expressions for spin-restricted closed-shell calculations. For four fixed spatial orbitals referred to as $p$, $q$, $r$, $s$, the Bethe-Salpeter kernel has the following spin structure
\begin{equation}
\left(
\begin{array}{cccc}
\Xi_{p\uparrow q\uparrow, r\uparrow s\uparrow} & \Xi_{p\uparrow q\uparrow, r \downarrow s \downarrow} & 0 & 0 \\ 
\Xi_{p\downarrow q\downarrow, r\uparrow s\uparrow} & \Xi_{p\downarrow q\downarrow, r \downarrow s\downarrow} & 0 & 0 \\ 
0 & 0 & \Xi_{p\uparrow q \downarrow, r \uparrow s\downarrow} & \Xi_{p\uparrow q\downarrow, r\downarrow s \uparrow} \\ 
0 & 0 & \Xi_{p \downarrow q\uparrow, r \uparrow s \downarrow} & \Xi_{p\downarrow q \uparrow, r \downarrow s \uparrow}
\end{array} \right),
\end{equation}
which can be brought to a diagonal form after rotation (see, e.g., Refs.~\onlinecite{TouZhuAngSav-PRA-10,TouZhuSavJanAng-JCP-11,AngLiuTouJan-JCTC-11})
\begin{equation}
\left( \begin{array}{cccc}
^1\Xi_{pq,rs} & 0 & 0 & 0 \\ 
0 & ^3\Xi_{pq,rs} & 0 & 0 \\ 
0 & 0 & ^3\Xi_{pq,rs} & 0 \\ 
0 & 0 & 0 & ^3\Xi_{pq,rs}
\end{array} \right),
\end{equation}
with a spin-singlet term $^1\Xi_{pq,rs} = 2 v_{pq,rs} - W_{pr,qs}$ and three degenerate spin-triplet terms $^3\Xi_{pq,rs} = - W_{pr,qs}$. It has been used that the Coulomb interaction $v$ and the screened interaction $W$ are spin independent: $v_{pq,rs} = v_{p\uparrow q\uparrow, r\uparrow s\uparrow}=v_{p\uparrow q\uparrow, r \downarrow s \downarrow}=v_{p\downarrow q\downarrow, r\uparrow s\uparrow}=v_{p\downarrow q\downarrow, r \downarrow s\downarrow}$ and $W_{pq,rs} = W_{p\uparrow q\uparrow, r\uparrow s\uparrow}=W_{p\uparrow q\uparrow, r \downarrow s \downarrow}=W_{p\downarrow q\downarrow, r\uparrow s\uparrow}=W_{p\downarrow q\downarrow, r \downarrow s\downarrow}$. The spin-adapted screened interaction is obtained by
\begin{equation}
W_{pq,rs} = \sum_{tu}  {^1}\epsilon^{-1}_{pq,tu}v_{tu,rs} \;,
\label{eq:W_spin}
\end{equation}
where $t$ and $u$ refer to spatial orbitals, and the singlet dielectric function ${^1}\epsilon_{pq,rs} = \epsilon_{p\uparrow q\uparrow,r\uparrow s\uparrow} + \epsilon_{p\uparrow q\uparrow ,r\downarrow s\downarrow}$ is given by
\begin{equation}
{^1}\epsilon_{pq,rs} = \delta_{pr}\delta_{qs} -  2 v_{pq,rs} \left[\chi_{0}(\omega=0)\right]_{rs,rs}.
\label{eq:singlet_eps}
\end{equation}
The bottom line is that the linear-response eigenvalue equation~(\ref{nonHermeq}) fully decouples into a singlet eigenvalue equation
\begin{equation}
\left(
\begin{array}{cc}
{^1}\bm{A} & {^1}\bm{B} \\
{^1}\bm{B}^* & {^1}\bm{A}^*
\end{array}
\right) 
\left(
\begin{array}{c}
^1\bm{X}_n \\
^1\bm{Y}_n\\
\end{array}
\right) 
= 
{^1\omega_n}
\left(
\begin{array}{cc}
\bm{1} & \bm{0} \\
\bm{0} & \bm{-1}
\end{array}
\right) 
\left(
\begin{array}{c}
^1\bm{X}_n \\
^1\bm{Y}_n\\
\end{array}
\right),
\label{nonHermeq1}
\end{equation}
with the matrices
\begin{subequations}
\begin{eqnarray}
^1A_{ia,jb} = \Delta \varepsilon_{ia,jb} +  2 v_{ia,jb} - W_{ij,ab},
\end{eqnarray}
\begin{eqnarray}
^1B_{ia,jb} = 2 v_{ia,bj} - W_{ib,aj},
\end{eqnarray}
\label{}
\end{subequations}
and a triplet eigenvalue equation
\begin{equation}
\left(
\begin{array}{cc}
{^3}\bm{A} & {^3}\bm{B} \\
{^3}\bm{B}^* & {^3}\bm{A}^*
\end{array}
\right) 
\left(
\begin{array}{c}
^3\bm{X}_n \\
^3\bm{Y}_n\\
\end{array}
\right) 
= 
{^3\omega_n}
\left(
\begin{array}{cc}
\bm{1} & \bm{0} \\
\bm{0} & \bm{-1}
\end{array}
\right) 
\left(
\begin{array}{c}
^3\bm{X}_n \\
^3\bm{Y}_n\\
\end{array}
\right),
\label{nonHermeq3}
\end{equation}
with the matrices
\begin{subequations}
\begin{eqnarray}
^3A_{ia,jb} = \Delta \varepsilon_{ia,jb} - W_{ij,ab},
\end{eqnarray}
\begin{eqnarray}
^3B_{ia,jb} = - W_{ib,aj}.
\end{eqnarray}
\label{}
\end{subequations}

\section{Example of H$_2$ in a minimal basis}
\label{sec:h2}
As a pedagogical example, we apply the BSE-GW method to the calculation of the excitation energies of H$_2$ in a minimal basis consisting of two Slater basis functions, $\varphi_a$ and $\varphi_b$, centered on each hydrogen atom and with the same exponent $\zeta=1$. This is a closed-shell molecule, therefore all the calculations are done with spin adaptation in a spatial orbital basis. The molecular orbitals are $\psi_1 = (\varphi_a + \varphi_b)/\sqrt{2(1+S_{ab})}$ (symmetry $\sigma_g$) and $\psi_2 = (\varphi_a - \varphi_b)/\sqrt{2(1-S_{ab})}$ (symmetry $\sigma_u$) where $S_{ab}$ is the overlap between $\varphi_a$ and $\varphi_b$. The matrix representations of all two-electron quantities in the space of spatial-orbital products are of the following form
\begin{equation}
\bm{P} = \left(
\begin{array}{cc|cc}
P_{11,11} & P_{11,22} & P_{11,12} & P_{11,21} \\
P_{22,11} & P_{22,22} & P_{22,12} & P_{22,21} \\
\hline
P_{12,11} & P_{12,22} & P_{12,12} & P_{12,21} \\
P_{21,11} & P_{21,22} & P_{21,12} & P_{21,21} \\
\end{array}
\right),
\end{equation}
and we refer to the upper left block as the (oo,vv) block, and to the bottom right block as the (ov,vo) block. All the values of the integrals as a function of the internuclear distance $R$ can be found in Ref.~\onlinecite{DewKel-JCE-71}. Note that, in the condensed-matter physics literature, a simplified version of H$_2$ in a minimal basis with only on-site Coulomb interaction is often used under the name ``half-filled two-site Hubbard model'' (see, e.g., Refs.~\onlinecite{AryGunRub-EPL-02,RomGuyRei-JCP-09})~\cite{RebTouSav-JJJ-XX-note}.

\subsection{BSE-GW method using the non-interacting Green's function}
\label{sec:BSE-GW-G0}

The simplest approximation in the BSE-GW method is to start from the non-interacting HF Green's function $G_0$, leading to the non-interacting HF linear response function $\chi_{IP}=-iG_0 G_0 = \chi_0$ whose matrix representation reads
\begin{equation}
\bm{\chi}_0(\omega)  = \left(
\begin{array}{cc|cc}
0&0&0&0 \\
0&0&0&0 \\
\hline
0&0&\dfrac{1}{\omega - \Delta \varepsilon} & 0 \\ 
0&0&0 & \dfrac{-1}{\omega + \Delta \varepsilon} \\
\end{array} 
\right),
\end{equation}
where $\Delta \varepsilon = \varepsilon_2 - \varepsilon_1$ is the difference between the energies of the molecular orbitals $\psi_2$ and $\psi_1$. The non-interacting linear response function has non-vanishing matrix elements only in the (ov,vo) block, but it will be necessary to consider the other blocks as well for the screened interaction $W$. The matrix of the Coulomb interaction is
\begin{equation}
\mb{v} = 
\left(
\begin{array}{cc|cc}
J_{11} & J_{12} & 0 & 0 \\
J_{12} & J_{22} & 0 & 0 \\
\hline
0 & 0 & K_{12} & K_{12} \\ 
0 & 0 & K_{12} & K_{12}
\end{array} 
\right),
\label{vmat}
\end{equation}
where $J_{pq}=\langle pq|pq\rangle$ and $K_{pq}=\langle pq|qp\rangle$ are the usual Coulomb and exchange two-electron integrals over the molecular orbitals $\psi_1$ and $\psi_2$. The off-diagonal blocks of $v$ are zero by symmetry for H$_2$ in a minimal basis, but this is not the case in general.
By matrix product and inversion, we get the static singlet dielectric matrix
\begin{equation}
^1\bm{\epsilon} = \left(
\begin{array}{cc|cc}
1 & 0 & 0 & 0 \\
0 & 1 & 0 & 0 \\
\hline
0 & 0 & 1+\dfrac{2K_{12}}{ \Delta \varepsilon} & \dfrac{2K_{12}}{\Delta \varepsilon} \\ 
0 & 0 & \dfrac{2K_{12}}{ \Delta \varepsilon} & 1+\dfrac{2K_{12}}{\Delta \varepsilon}
\end{array} 
\right),
\end{equation}
which, in this case, is block diagonal with the (oo,vv) block being the identity. By using its inverse, we finally get the static screened interaction matrix
\begin{equation}
\mb{W} = 
\left(
\begin{array}{cc|cc}
J_{11} & J_{12} & 0 & 0 \\
J_{12} & J_{22} & 0 & 0 \\
\hline
0 & 0 &
\dfrac{K_{12} }{1 + 4K_{12}/\Delta\varepsilon} &  \dfrac{K_{12} }{1 + 4K_{12}/\Delta\varepsilon}\\ 
0 & 0 &
\dfrac{K_{12} }{1 + 4K_{12}/\Delta\varepsilon} & \dfrac{K_{12} }{1 + 4K_{12}/\Delta\varepsilon}
\end{array} 
\right),
\end{equation}
which is block diagonal and the (oo,vv) block is just the bare Coulomb interaction in the case of H$_2$ in a minimal basis, but this is not generally true. We have then everything to construct the $^1A$ and $^1B$ matrices of equation~(\ref{nonHermeq1}) for singlet excitations, which in the present case are just one-dimensional 
\begin{subequations}
\begin{eqnarray}
^1A = \Delta \varepsilon +  2 K_{12} - J_{12},
\end{eqnarray}
\begin{eqnarray}
^1B = 2K_{12} - \dfrac{K_{12} }{1 + 4K_{12}/\Delta\varepsilon}.
\end{eqnarray}
\label{}
\end{subequations}
and the $^3A$ and $^3B$ matrices of equation~(\ref{nonHermeq3}) for triplet excitations
\begin{subequations}
\begin{eqnarray}
^3A = \Delta \varepsilon  - J_{12},
\end{eqnarray}
\begin{eqnarray}
^3B = - \dfrac{K_{12} }{1 + 4K_{12}/\Delta\varepsilon}.
\end{eqnarray}
\label{}
\end{subequations}
Solving then the response equations by the standard Casida approach~\cite{Cas-INC-95}, we get the singlet excitation energy
\begin{equation}
\begin{split}
^1\omega = & \left[ 
\left( \Delta\varepsilon + 4 K_{12} - J_{12} - \dfrac{K_{12} }{1 + 4K_{12}/\Delta\varepsilon}\right) \right. \\
& \times \left. \left( \Delta  \varepsilon - J_{12} + \dfrac{K_{12} }{1 + 4K_{12}/\Delta\varepsilon} \right) \right]^{1/2}
\end{split},
\label{1omega}
\end{equation}
and the triplet excitation energy
\begin{equation}
\begin{split}
^3\omega =& \left[
\left( \Delta\varepsilon - J_{12}- \dfrac{K_{12} }{1 + 4K_{12}/\Delta\varepsilon}\right)\right. \\
& \times \left. \left( \Delta  \varepsilon - J_{12} + \dfrac{K_{12} }{1 + 4K_{12}/\Delta\varepsilon} \right) \right]^{1/2}.
\end{split}
\label{3omega}
\end{equation}
Note that, for this simple system, the $A$ terms have the usual TDHF or configuration interaction singles (CIS) forms, and the screening has an effect only on the $B$ terms, decreasing the exchange integral $K_{12}$ by a factor of $1 + 4K_{12}/\Delta\varepsilon$. Therefore, in the Tamm-Dancoff approximation~\cite{HirHea-CPL-99}, which consists in neglecting $B$, the effect of screening would be lost and the method would be equivalent to CIS. It is interesting to analyze the effect of the screening as a function of the internuclear distance $R$. For small $R$, the orbital energy difference $\Delta\varepsilon$ is much greater than the exchange integral $K_{12}$, so the screening factor $1 + 4K_{12}/\Delta\varepsilon$ is close to $1$ and TDHF excitation energies are recovered. For large $R$ (dissociation limit), $\Delta\varepsilon$ goes to zero, so the screening factor diverges and the term $K_{12}/(1 + 4K_{12}/\Delta\varepsilon)$ vanishes.

\begin{figure*}
\includegraphics[width=0.45\textwidth]{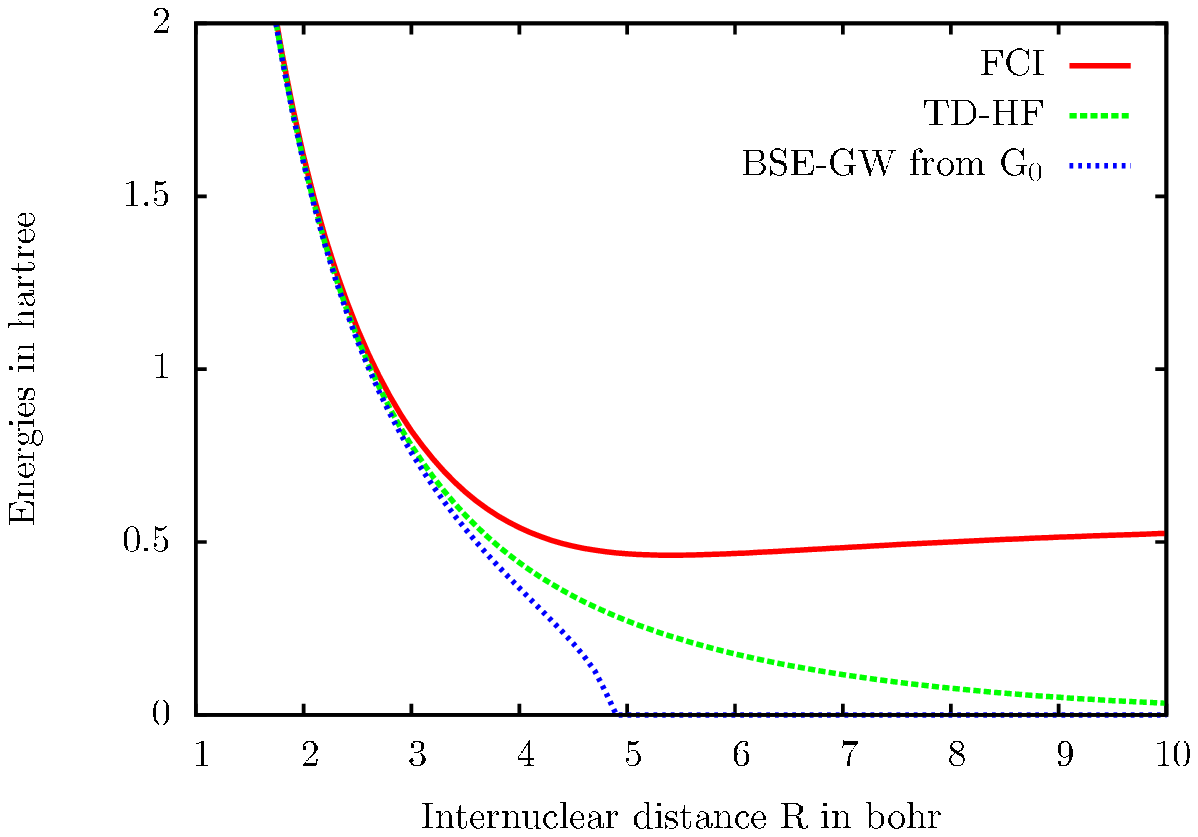}
\includegraphics[width=0.45\textwidth]{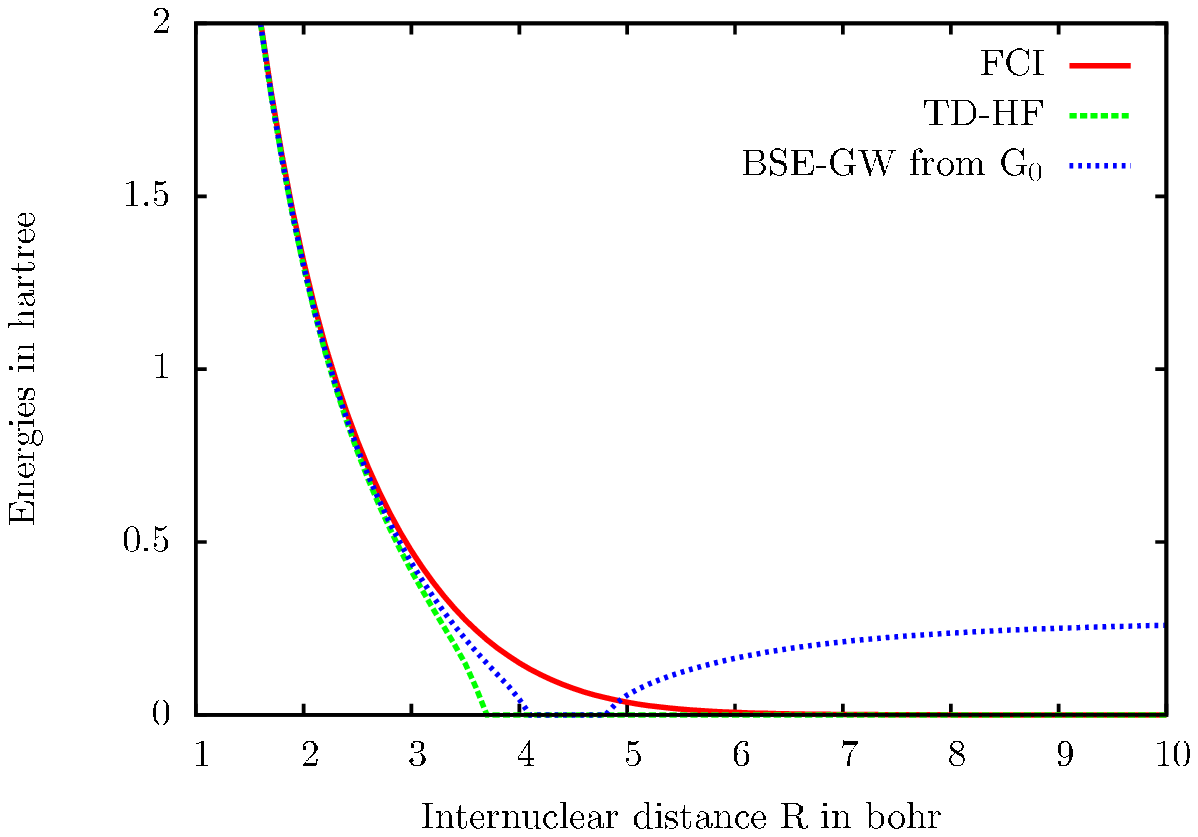}
\caption{Excitation energies of the singlet $^1\Sigma_u^+$ (left) and the triplet $^3\Sigma_u^+$ (right) states of H$_2$ in a minimal basis as a function of the internuclear distance $R$ calculated by FCI, TDHF, and BSE-GW with the non-interacting HF Green's function $G_0$.}
\label{fig:H2G0W}
\end{figure*}

The excitation energies from the ground state $^1\Sigma_g^+$ to the first singlet $^1\Sigma_u^+$ and triplet $^3\Sigma_u^+$ excited states are plotted as a function of $R$ in Figure~\ref{fig:H2G0W}. The reference curves are from a full configuration-interaction (FCI) calculation giving the exact excitation energies in this basis. In a minimal basis, the singlet $^1\Sigma_u^+$ excited state is constrained to dissociate into the ionic configuration $H^-...~H^+$, and so in the dissociation limit $R\to\infty$ the exact singlet excitation energy goes to a constant, $I-A\approx 0.625$ hartree where $I$ and $A$ are the ionization energy and electron affinity of the hydrogen atom. The triplet $^3\Sigma_u^+$ dissociates into the neutral configuration $H^\bullet...~H^\bullet$, as does the ground state, and so the exact triplet excitation energy goes to zero in the dissociation limit. TDHF gives accurate excitation energies for small $R$, but gives qualitatively wrong curves in the dissociation limit. For the singlet state, the TDHF excitation energy goes to zero, a wrong behavior inherited from the vanishing $\Delta\varepsilon$ in this limit. For the triplet state, the TDHF response equation suffers from a triplet instability for $R\geq 4$ bohr and the excitation energy becomes imaginary. It is known that TDDFT with standard density-functional approximations gives similarly incorrect energy curves~\cite{CaiRei-JCP-00,CasGutGuaGadSalDau-JCP-00,GriGisGorBae-JCP-00,AryGunRub-EPL-02,GieBae-CPL-08}. The BSE-GW method using the non-interacting HF Green's function $G_0$ gives accurate excitation energies at small $R$, but fails in the dissociation limit. The singlet excitation energy becomes imaginary for $R\geq 4.9$ bohr. Indeed, in the dissociation limit, $\Delta\varepsilon$ goes to zero and equation~(\ref{1omega}) leads to a negative term under the square root: $^1\omega \to \sqrt{(4K_{12}-J_{12})(-J_{12})}$. Similarly, the BSE-GW triplet excitation energy is imaginary between $R=4.0$ and $R=4.9$ bohr, and incorrectly tends to a non-zero value in the dissociation limit.

The BSE-GW method using the non-interacting HF Green's function $G_0$ thus badly fails for H$_2$ in the dissociation limit. As this method is based on a single-determinant reference, this should not come as a surprise. However, the BSE approach also allows one to start from an interacting Green's function $G$ taking into account the multiconfigurational character of stretched H$_2$. We will now test this alternative approach.

\subsection{BSE-GW method using the exact Green's function}
\label{sec:BSE-GW-Gexact}

\subsubsection{Independent-particle response function}

\begin{figure*}
\centering
\includegraphics{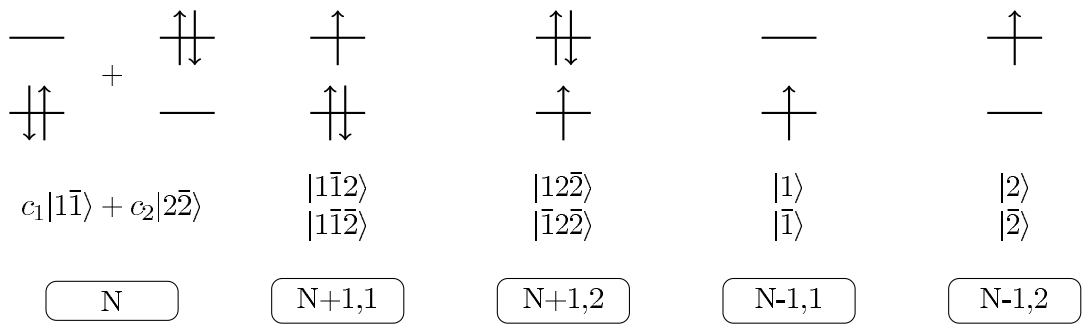}
\caption{$N$-electron ground state, and  $(N \pm 1)$-electron states for H$_2$ in a minimal basis. \label{fig:conf_H2}}
\end{figure*}

We apply the BSE-GW equations~(\ref{BSEGW})-(\ref{epsilonstatic}) with the independent-particle response function $\chi_{IP}=-i GG$ constructed from the exact one-particle Green's function $G$, and which can be calculated by the Lehmann formula~(\ref{ChiIPLehmann}) using the $N$-electron ground state and the $(N \pm 1)$-electron states. The states to consider for H$_2$ in a minimal basis are given in Figure~\ref{fig:conf_H2}. The ground state is composed of two Slater determinants, its energy is $E_N = 2\varepsilon_1 -J_{11} + E_c$ where $E_c= \Delta - \sqrt{\Delta^2 +K_{12}^2}$ is the correlation energy with $2\Delta = 2\Delta \varepsilon + J_{11} + J_{22}-4J_{12}+2K_{12}$. The coefficients of the determinants are determined by $c_2 = c_1 K_{12}/(\Delta + \sqrt{K_{12}^2+\Delta^2})$ and $c_1^2 +c_2^2=1$. The energies of the two $(N+1)$-electron states are: $E_{N+1,1}  = 2 \varepsilon_1 + \varepsilon_2 - J_{11} $ and $E_{N+1,2} = 2\varepsilon_2 + \varepsilon_1 -J_{11} +J_{22} -2J_{12} + K_{12}$. The energies of the two $(N-1)$-electron states are: $E_{N-1,1} = \varepsilon_1 -J_{11}$ and $E_{N-1,2} = \varepsilon_2-2J_{12} + K_{12}$. We thus obtain four poles for the exact one-particle Green's function. Two of them correspond to minus the electron affinities,
\begin{subequations}
\begin{equation}
{\cal E}_{2} = E_{N+1,1}-E_N = \varepsilon_2 -E_c,
\end{equation}
\begin{equation}
{\cal E}_{2}' = E_{N+1,2}-E_N = 2\varepsilon_2 - \varepsilon_1 +J_{22} -2J_{12} + K_{12}  - E_c,
\end{equation}
\end{subequations}
and the other two correspond to minus the ionization energies,
\begin{subequations}
\begin{equation}
{\cal E}_{1} = E_N-E_{N-1,1} = \varepsilon_1 +E_c,
\end{equation}
\begin{equation}
{\cal E}_{1}' = E_N-E_{N-1,2} = 2\varepsilon_1 - \varepsilon_2 -J_{11} + 2J_{12} - K_{12} + E_c. 
\end{equation}
\end{subequations}
In condensed-matter physics, ${\cal E}_{1}$ and ${\cal E}_{2}$ are associated with ``quasi-particle'' peaks of photoelectron spectra, while ${\cal E}_{1}'$ and ${\cal E}_{2}'$ are associated with ``satellites''.
The Dyson orbitals are also easily calculated, and we finally arrive at the matrix representation of $\chi_{IP}$ in the basis of the products of spatial orbitals
\begin{equation}
\bm{\chi}_{IP}(\omega) =
\left(
\begin{array}{cc|cc}
\chi_{IP,11}(\omega) & 0 & 0 & 0 \\
0 & \chi_{IP,22}(\omega) & 0 & 0 \\
\hline
0 & 0 & \chi_{IP,12}(\omega) & 0 \\
0 & 0 & 0 & \chi_{IP,21}(\omega) \\
\end{array}
\right),
\end{equation}
with the matrix elements
\begin{subequations}
\begin{equation}
\chi_{IP,11}(\omega) = \dfrac{ c_1^2c_2^2 }{\omega - ({\cal E}_{2}'-{\cal E}_{1})}  - \dfrac{c_1^2c_2^2 }{\omega + ({\cal E}_{2}'-{\cal E}_{1})},
\end{equation}
\begin{equation}
\chi_{IP,22}(\omega) = \dfrac{ c_1^2c_2^2 }{\omega - ({\cal E}_{2}-{\cal E}_{1}')}- \dfrac{c_1^2c_2^2 }{\omega + ({\cal E}_{2}-{\cal E}_{1}')},
\end{equation}
\begin{equation}
\chi_{IP,12}(\omega) = \dfrac{c_1^4}{\omega - ({\cal E}_{2}-{\cal E}_{1})} -\dfrac{c_2^4 }{\omega + ({\cal E}_{2}'-{\cal E}_{1}')},
\end{equation}
\begin{equation}
\chi_{IP,21}(\omega) = \dfrac{c_2^4}{\omega - ({\cal E}_{2}'-{\cal E}_{1}')}  -\dfrac{c_1^4 }{\omega + ({\cal E}_{2}-{\cal E}_{1})}.
\end{equation}
\end{subequations}

\begin{figure}[t]
\centering
\includegraphics[width=0.45\textwidth]{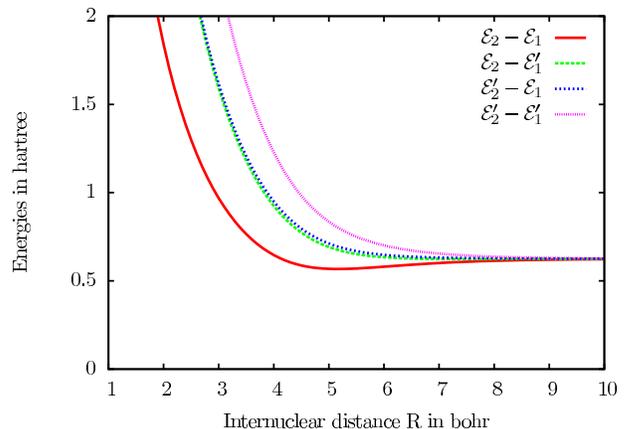}
\caption{Positive poles of the independent-particle linear response function in function of the internuclear distance R}
\label{fig:chi_ip_pole}
\end{figure}

Therefore, whereas ${\chi}_{0}(\omega)$ has only one positive pole, ${\chi}_{IP}(\omega)$ has four distinct positive poles (and four symmetric negative poles). These poles are plotted in Figure~\ref{fig:chi_ip_pole}. The lowest one, ${\cal E}_{2}-{\cal E}_{1}$, called fundamental gap in the condensed-matter physics literature, can be considered as an approximation to a neutral single excitation energy since in the limit of non-interacting particles it equals the difference of the orbital eigenvalues $\Delta \varepsilon = \varepsilon_2 - \varepsilon_1$. The two intermediate poles, ${\cal E}_{2}'-{\cal E}_{1}$ and ${\cal E}_{2}-{\cal E}_{1}'$, can be interpreted as approximations to a double excitation energy since they reduce to $2\Delta \varepsilon$ in the limit of non-interacting particles. Surprisingly, the highest pole, ${\cal E}_{2}'-{\cal E}_{1}'$, reduces to $3\Delta \varepsilon$ in this limit and it is thus tempting to associate it with a triple excitation even though the system contains only two electrons! In the dissociation limit $R\to\infty$, the four poles tends to the same value, i.e. $I-A \approx 0.625$ hartree which is also minus twice the correlation energy $-2E_c$, showing that the non-vanishing fundamental gap in this limit is a correlation effect. Note that it has been shown~\cite{RomGuyRei-JCP-09} that the non-self-consistent GW approximation ($G_0 W_0$) to the one-particle Green's function gives a fundamental gap which is too small by a factor of $2$ in the dissociation limit,  so we do not consider this approximation here.

\subsubsection{Excitation energies}

\begin{figure}[t]
\centering
\includegraphics[width=0.45\textwidth]{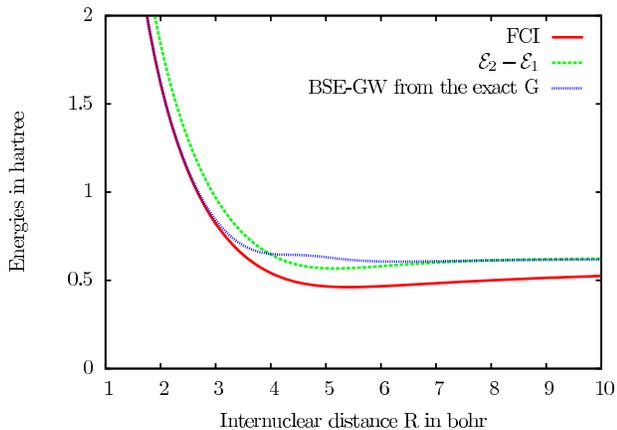}
\caption{Excitation energy of the singlet $^1\Sigma_u^+$ state of H$_2$ in a minimal basis as a function of the internuclear distance $R$ calculated by FCI and BSE-GW with the exact Green's function. The lowest pole of ${\chi}_{IP}(\omega)$, the fundamental gap ${\cal E}_{2}-{\cal E}_{1}$, is also plotted for comparison.}
\label{fig:h2_gw_sing_ovvo}
\end{figure}

Having calculated the independent-particle response function, the next steps of the BSE-GW calculation of the excitation energies proceed similarly as in Section~\ref{sec:BSE-GW-G0}, even though the expressions get more complicated. From the matrix $\bm{\chi}_{IP}(\omega=0)$ and the Coulomb interaction matrix~(\ref{vmat}), we calculate the singlet dielectric matrix which is still block diagonal but the upper left block is no longer the identity matrix. We calculate then the static screened interaction matrix which is still block diagonal but the elements of its upper left block are now also affected by screening. We can then construct the corresponding singlet and triplet Bethe-Salpeter kernel ${^1}\bm{\Xi}$ and ${^3}\bm{\Xi}$. The response eigenvalue equations~(\ref{nonHermeq1}) and~(\ref{nonHermeq3}) are no longer applicable, so the singlet excitation energies are found by searching the values of $\omega$ giving vanishing eigenvalues of the inverse singlet linear response matrix ${^1}\bm{\chi}(\omega)^{-1} = \bm{\chi}_{IP}(\omega)^{-1} - {^1}\bm{\Xi}$, and the triplet excitation energies are found by searching the values of $\omega$ giving vanishing eigenvalues of the inverse triplet linear response matrix ${^3}\bm{\chi}(\omega)^{-1} = \bm{\chi}_{IP}(\omega)^{-1} - {^3}\bm{\Xi}$. For H$_2$ in a minimal basis, ${^1}\bm{\chi}(\omega)^{-1}$ and ${^3}\bm{\chi}(\omega)^{-1}$ are $4\times4$ matrices which are block diagonal, the (oo,vv) block being uncoupled to the (ov,vo) block. For both the singlet and triplet cases, the four positive poles of ${\chi}_{IP}(\omega)$ transform into four excitation energies (plus four symmetric de-excitation energies). 

Among the two positive excitation energies coming from the (ov,vo) block of the matrix ${^1}\bm{\chi}(\omega)^{-1}$, the lowest one is identified with the first singlet $^1\Sigma_u^+$ excitation energy, which is called the optical gap. It is plotted in Figure~\ref{fig:h2_gw_sing_ovvo} and compared with the reference FCI excitation energy and also with the fundamental gap ${\cal E}_{2}-{\cal E}_{1}$ to highlight the effect of the Bethe-Salpeter kernel. At small internuclear distance, $R \leq 3$ bohr, the Bethe-Salpeter kernel brings the BSE-GW curve is very close to the FCI curve. For large $R$, the BSE-GW excitation energy follows the curve of the fundamental gap, which slightly overestimates the excitation energy at $R=10$ bohr but eventually goes to the correct limit $I-A$ when $R\to\infty$. Thus, contrary to the BSE-GW method using the non-interacting Green's function, the obtained excitation energy curve has now a correct shape. This relies on the fundamental gap being a good starting approximation to the optical gap. As regards the second excitation energy coming from the (ov,vo) block of the matrix ${^1}\bm{\chi}(\omega)^{-1}$ which is connected to highest pole ${\cal E}_{2}'-{\cal E}_{1}'$ of ${\chi}_{IP}(\omega)$, it is a spurious excitation due to the approximate Bethe-Salpeter kernel used.

\begin{figure}[t]
\centering
\includegraphics[width=0.45\textwidth]{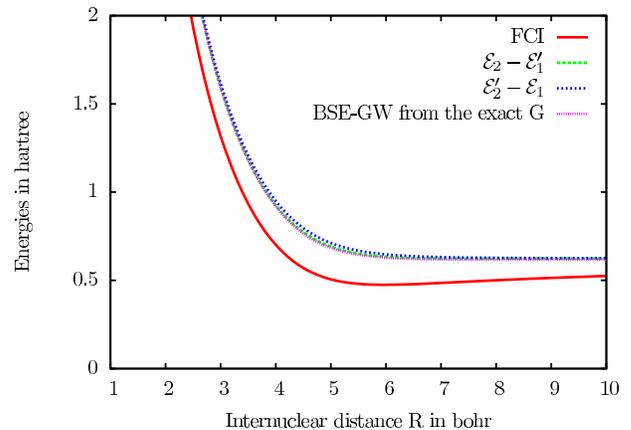}
\caption{Excitation energy of the second singlet $^1\Sigma_g^+$ state of H$_2$ in a minimal basis as a function of the internuclear distance $R$ calculated by FCI and BSE-GW with the exact Green's function. The poles ${\cal E}_{2}'-{\cal E}_{1}$ and ${\cal E}_{2}-{\cal E}_{1}'$ of ${\chi}_{IP}(\omega)$ are also plotted for comparison.}
\label{h2_gw_sing_oovv}
\end{figure}

The lowest positive excitation energy coming from the (oo,vv) block of the matrix ${^1}\bm{\chi}(\omega)^{-1}$ is identified with the second singlet $^1\Sigma_g^+$ excited state which has a double excitation character. It is plotted in Figure~\ref{h2_gw_sing_oovv} and compared with the FCI excitation energy for this state and with the poles ${\cal E}_{2}'-{\cal E}_{1}$ and ${\cal E}_{2}-{\cal E}_{1}'$ of ${\chi}_{IP}(\omega)$. It is noteworthy that the BSE-GW method starting from ${\chi}_{IP}(\omega)$ instead of ${\chi}_{0}(\omega)$ but using a frequency-independent kernel does describe this double-excitation state with an overall correct shape for the energy curve. However, the BSE-GW excitation energy is almost identical to the two poles ${\cal E}_{2}'-{\cal E}_{1}$ and ${\cal E}_{2}-{\cal E}_{1}'$. The Bethe-Salpeter kernel in the static GW approximation thus brings virtually no improvement for this state over the starting poles of ${\chi}_{IP}(\omega)$. The (oo,vv) block of the matrix ${^1}\bm{\chi}(\omega)^{-1}$ also gives a second higher excitation energy that is spurious.

\begin{figure}[t]
\centering
\includegraphics[width=0.45\textwidth]{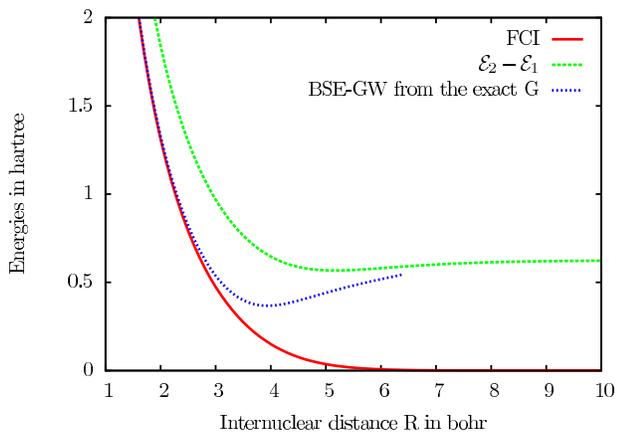}
\caption{Excitation energy of the triplet $^3\Sigma_u^+$ state of H$_2$ in a minimal basis as a function of the internuclear distance $R$ calculated by FCI and BSE-GW with the exact Green's function. The lowest pole of ${\chi}_{IP}(\omega)$, the fundamental gap ${\cal E}_{2}-{\cal E}_{1}$, is also plotted for comparison.}
\label{h2_gw_trip_ovvo}
\end{figure}

We finally consider the triplet excited state $^3\Sigma_u^+$. The lowest positive excitation energy coming from the (ov,vo) block of the matrix ${^3}\bm{\chi}(\omega)^{-1}$ should be identified with this state. It is plotted in Figure~\ref{h2_gw_trip_ovvo} and compared with the FCI excitation energy for this state and with the fundamental gap ${\cal E}_{2}-{\cal E}_{1}$. For small internuclear distances, $R\leq 3$ bohr, the BSE-GW method gives an accurate excitation energy, but for larger $R$, instead of going to zero, the BSE-GW excitation energy follows the fundamental gap until the excitation energy becomes imaginary for $R\geq 6.5$ bohr. The problem is that the poles of ${\chi}_{IP}(\omega)$ are the same for both the singlet and triplet cases, and the fundamental gap ${\cal E}_{2}-{\cal E}_{1}$ is not a good starting approximation to the triplet excitation energy in the dissociation limit. The Bethe-Salpeter kernel in the static GW approximation is not able of compensating for this bad starting point. In addition to this excitation energy, the BSE-GW method gives three other spurious triplet excitation energies.

\section{Conclusion}
\label{sec:conclusion}

We have applied the BSE approach in the static GW approximation for the calculation of the excitation energies on the toy model of H$_2$ in a minimal basis. We have tested two variants for the starting one-particle Green's function: the non-interacting HF one and the exact one. Around the equilibrium internuclear distance, both variants give accurate excitation energies to the first singlet $^1\Sigma_u^+$ and triplet $^3\Sigma_u^+$ excited states. In the dissociation limit, however, the two variants differ. The first variant, starting from the non-interacting one-particle Green's function, badly fails in this limit for both the singlet and triplet states, giving imaginary excitation energies. The second variant, starting from the exact one-particle Green's function, gives a qualitatively correct energy curve for the singlet $^1\Sigma_u^+$ excited state up to the dissociation limit. This relies on the fact that the fundamental gap (given by the one-particle Green's function) is a good starting approximation to the first singlet excitation energy. However, the same variant gives an incorrect energy curve for the triplet $^3\Sigma_u^+$ excited state in the dissociation limit. In this case, the fundamental gap is a bad starting approximation to the first triplet excitation energy.

The second BSE variant using the exact one-particle Green's function gives more excitation energies that the first BSE variant. Most of them are spurious excitations due to the approximate Bethe-Salpeter kernel used. However, one of them can be identified with the excitation energy to the singlet $^1\Sigma_g^+$ excited state which has a double excitation character. It is remarkable that such a double excitation can be described at all without using a frequency-dependent kernel. However, the Bethe-Salpeter kernel in the static GW approximation is insufficient to describe accurately the energy curve of this state, even around the equilibrium distance.


\section*{Acknowledgments}
We thank L. Reining and F. Sottile (\'Ecole Polytechnique, Palaiseau, France), and K. Pernal (Politechnika \L\'odzka, \L\'od\'z, Poland) for discussions.

\bibliographystyle{apsrev}

\begin{thebibliography}{47}
\bibitem[{Reb()}]{RebTouSav-JJJ-XX-note}
\bibinfo{note}{With the notations used here, the Hubbard model is obtained for
  $\Delta\varepsilon = 2 t$ and $J_{11}=J_{22}=J_{12}=J_{12}=K_{12}=U/2$ where
  $t$ is the hopping parameter and $U$ is the on-site Coulomb interaction.}
\end{thebibliography}

\end{document}